# Un discurso inédito de Einstein en su visita a la Argentina en 1925

*Alejandro Gangui, IAFE/Conicet,
CEFIEC. FCEyN-UBA, Buenos Aires*
y
*Eduardo L. Ortiz
Imperial College, Londres*

## Introducción

En un artículo publicado por Mauricio Nirenstein en 1925, pocos meses después de la visita de Albert Einstein a la Argentina, el autor hizo interesantes comentarios y dio diversas referencias relativas a esa visita. En particular, Nirenstein hizo referencia a una conversación sostenida con Einstein, en la que éste último le habría hecho reflexiones acerca de su visión de la epistemología de las ciencias físicas. Si bien los otros extractos de Nirenstein no dejan de tener interés, el que se refiere al aspecto epistemológico presenta facetas que vale la pena discutir en cierto detalle.

En una nota a ese artículo se indica que Einstein habría facilitado a Nirenstein notas de un discurso que pensaba pronunciar en Buenos Aires. Y así, a partir de 1931, aparecieron en revistas de Buenos Aires versiones del que fue llamado "discurso inédito" de Einstein.

En la primera referencia, la de 1931, indicaba que ese artículo habría sido elaborado por Einstein durante su viaje, en vísperas de su llegada a Buenos Aires, explicando que había sido pensado como conferencia de introducción a su curso sobre la teoría de la relatividad, y que había sido descartado a último momento en favor de un enfoque más estricto, ciñéndo su curso a temas específicos de física teórica. Se decía también que el texto original, en idioma alemán, había quedado en manos de Nirenstein.

En este trabajo vamos a discutir estos hechos y personajes. Haremos un análisis breve del "discurso inédito" y lo compararemos con las dos conferencias de introducción a su curso sobre la teoría de la relatividad, dictadas, respectivamente, ante la Universidad de Buenos Aires y, al día siguiente, en la Facultad de Ciencias Exactas.

## Mauricio Nirenstein

A comienzos de su carrera, Mauricio Nirenstein mostró interés por las letras, publicando ensayos breves, cuentos y poesías. Esos intereses lo llevaron a vincularse con el movimiento literario del Buenos Aires de fines del siglo XIX, muy particularmente a los grupos centrados alrededor de la figura del naturalista y escritor Dr. Eduardo L. Holmberg.

Aunque la obra literaria original de Nirenstein fue limitada y se detuvo tempranamente, sus antiguos intereses literarios lo llevaron a profundizar en el estudio de la literatura de Europa septentrional, alcanzando luego la cátedra en esa especialidad en la Facultad de Filosofía y Letras de la Universidad de Buenos Aires.

Junto con su carrera docente, Nirenstein desarrolló una larga y exitosa carrera de gestión dentro de la UBA, manteniendo una relación estrecha con el cirujano José Arce, el filósofo Coriolano Alberini, y otros decanos y rectores de ese período. Algunos de ellos, representantes de corrientes de derecha conservadora dentro de la UBA, jugaron un papel importante en el período de la contra-Reforma. Nirenstein se incorporó a la secretaría de la UBA en 1897, siendo designado pro-Secretario en 1906 y, finalmente, Secretario en 1922, bajo el rectorado de Arce. Conservó ese cargo hasta su jubilación, en 1930.

Justamente en el año en el que Nirenstein se hizo cargo de la Secretaria de la UBA, esa universidad -principalmente por iniciativas de Leopoldo Lugones, del físico e ingeniero Jorge Duclout y del matemático Julio Rey Pastor (quien había sido un elemento importante en la visita de Einstein a España)- iniciaba gestiones tendientes a lograr que Einstein visitara la Argentina por espacio de un mes y dictara un ciclo de conferencias sobre su teoría de la relatividad. Esa visita se concretó tres años más tarde, cuando el eminente profesor hizo también visitas, mucho más breves, a las ciudades de Córdoba y de La Plata.

A principios de la década de 1920 Nirenstein participó también, activamente, dentro del grupo fundador de la Asociación Hebraica Argentina, más tarde designada con el nombre de Sociedad Hebraica Argentina (SHA). El grupo de la SHA buscaba contribuir a dar a la nueva generación de intelectuales judío-argentinos una mayor visibilidad en el horizonte cultural de esos años. Las gestiones para la invitación de Einstein a visitar la Argentina formaban también una parte central dentro del cuadro de iniciativas de esa asociación que, con razón, percibía a Einstein como el representante más destacado -y visible- en Europa de una nueva generación de científicos de origen judío, pacifistas y sostenedores de ideas sociales y políticas de avanzada.

Con contactos en ambos grupos, Nirenstein jugó un papel sino único, por lo menos destacado en la administración de la visita de Einstein. La UBA lo comisionó, en su carácter de Secretario de la misma, para integrar el grupo de universitarios argentinos que viajó a Montevideo a recibir al sabio y luego acompañarlo en el *Cap Polonio* en la etapa final de su viaje a Buenos Aires. Desde ese momento, Nirenstein se posicionó frente al sabio como uno de sus más estrechos asesores. De este modo Nirenstein tuvo una influencia considerable en el desarrollo de esa visita. Es posible argumentar que en alguna medida contribuyó a orientarla en un sentido que él, y posiblemente un grupo importante de quienes contribuyeron a auspiciarla -y también a financiarla- interpretaba como más en consonancia con los resultados que de ella esperaban.

Como hemos indicado en otra parte [Ortiz, 1995; Gangui y Ortiz, 2005], así lo sugiere el curioso cambio de la agenda amplia y de tono político y polémico propuesta inicialmente por Einstein para su relación con el gran público de Argentina. Así lo sugiere también la elección del tema del *Pan-Europeismo* (es decir, de la posibilidad de integrar la vida cultural y política de Europa de manera de hacer imposible la repetición de la Guerra Mundial que acababa de concluir) en la colaboración que envió al diario *La Prensa*, y que apareció al momento de su llegada a Buenos Aires (el 24 de Marzo, justo un día antes de su arribo en barco) [Einstein, 1925a.].

En la dedicatoria de una fotografía que obsequió a Nirenstein hacia la fecha de su partida de la Argentina, Einstein mismo aludió risueñamente al rol contemporizador que Nirenstein había jugado en su visita. En el *Diario íntimo* de su viaje por sudamérica, podemos leer la siguiente frase de Einstein:

> Bien conducido por su mano
> Caminé con coraje por esta tierra.
> Se supone que él sabe lo que se requiere
> Para asegurar que nadie se "enoje".
> Él sabe mirar en el corazón de cada uno.
> Este es el Señor Profesor Nirenstein.
> Mi gratitud a su alma indulgente
> Le sea transmitida con esta Foto del Sabio.

[déjennos terminar su estrofa en el original para que apreciemos su estilo]

> Allen sieht ins Herz hinein
> Herr Professor Nirenstein
> Dank an seine Seele mild
> Kuend'ihm dieses Sabio-Bild.

(tanto había oido Einstein que lo llamaban SABIO que acabó haciéndose una broma a él mismo, firmando con el nombre de "Sabio"...)

### El discurso inédito de Einstein: la versión dialogada de Nirenstein

Poco después de la partida de Einstein de la Argentina, en Septiembre de 1925, Nirenstein [Nirenstein, 1925] publicó un artículo en el tomo XVIII de *Verbum*, la revista del Centro de Estudiantes de Filosofía y Letras, titulado: *Einstein en Buenos Aires*.

En ese artículo hay interesantes referencias a varios aspectos extra-científicos de la visita de Einstein a la Argentina:

- los diferentes matices con los que Einstein fue recibido por diferentes sectores de la comunidad judía-argentina, tanto por asimilacionistas (grupo al que Nirenstein pertenecía) como por los más radicales sionistas. (Esto, en el apartado *La propaganda sionista*);
- sus impresiones sobre los medios de comunicación (en los apartados *Einstein y los diarios* y *El cinematógrafo*);
- también aparece una apología de la percepción que Einstein tenía del talento filosófico de Alberini (en *Einstein, la Academia de Ciencias de Prusia y Alberini*).
- Sigue luego una semblanza de la visita que Einstein hizo a la casa particular de Ducluot (quien, como él, había sido antiguo alumno del Politécnico de Zurich) y que se encontraba enfermo, titulada: *El ingeniero Duclout y la moral de Kant*.

Aunque un gestor científico y administrativo importante de la visita, para 1925 Duclout se encontraba seriamente enfermo; jugó en ella un papel relativamente limitado durante su desarrollo.

- La nota de Nirenstein finaliza con una sección (que se anuncia en una nota al pie que baja del título mismo del artículo, pero que no se incluyó en el texto) que debió haberse titulado: *Una disertación epistemológica*.

Esta sección final contiene una referencia aun más interesante. Se trata de la trascripción, posiblemente metafórica, de una conversación epistemológica que Nirenstein habría sostenido con Einstein en el automóvil, de regreso a Buenos Aires, luego de la visita a Duclout. Nirenstein nos explica que, "Como [Einstein] habla en alemán, su expresión es nítida y de precisión admirable" destacando un aspecto no desconocido de las dificultades de comunicación del sabio en la Argentina.

En esa conversación Einstein habría expuesto sus ideas acerca de la evolución de la física contemporánea. La nota que baja del título del artículo y a la que hemos aludido más atrás explica que esta parte final "tiene por base un escrito inédito de Einstein." Sin embargo, nada de eso se dice en el texto del artículo.

Una nueva referencia a ese artículo inédito de Einstein apareció en un artículo que Enrique Espinosa publicó en *La Nación* [Espinosa, 1934] comentando el nuevo libro de Einstein sobre su visión del mundo. En esa reseña Espinosa hizo referencia al "discurso inédito" difundido tres años antes por *La Vida Literaria*. Años más tarde, en una nota de Diego F. Pró insertada en el primer volumen de su edición del *Epistolario* de Alberini [Alberini, 1980], se hizo nuevamente referencia a que el discurso inédito fue escrito por Einstein a bordo del *Cap Polonio* en camino a Buenos Aires, y que fue dejado en manos de Nirenstein.

A través de las notas incluidas por Einstein en su *Diario íntimo* sabemos que durante su viaje en barco Einstein alternó preocupaciones científicas con lecturas filosóficas. Uno de los autores leídos a bordo del *Cap Polonio* fue Émile Meyerson (1859-1933), pensador anti-positivista y autor, en 1924, de la obra *La déduction relativiste* [Meyerson, 1924], a la que tres años más tarde Einstein mismo se refirió elogiosamente [Einstein, 1928].

Espinosa señala en 1934 que las notas de Einstein fueron publicadas en el número de Abril de 1931 de *La Vida Literaria* [Einstein, 1931], una revista dirigida por el ensayista, periodista y promotor literario Samuel Glusberg, impulsada por un grupo de destacados intelectuales de Buenos Aires de aquella época, y que contó con una lista amplia de colaboradores. Los ejemplares de esa revista son hoy muy raros. De acuerdo con la referencia dada por Espinosa y repetida en el *Epistolario*, la traducción de esa "conferencia inédita" cuyo texto, como hemos dicho, Einstein habría entregado a Nirenstein, había sido hecha por Baldomero Sanín Cano (1861-1957).

No debe sorprendernos que el escritor y diplomático colombiano residente en Buenos Aires Sanín Cano haya sido señalado como el autor de la traducción, ya que tenía buen dominio de varias lenguas, y había leído a diversos filósofos alemanes en su idioma original. Como Einstein, aunque más tarde, estuvo también vinculado al *Instituto Internacional de Cooperación Intelectual* de la *Sociedad de las Naciones*.

El artículo inédito continuó dando que hablar. Hacia fines de 1955, luego de la muerte del célebre científico, la revista *DAVAR*, órgano de la SHA, republicó esa nota con el mismo título, *Un discurso inédito de Einstein*, sin agregar nueva información acerca de ese discurso [Einstein, 1955] y una nota de Espinosa (que ahora firma Espinoza, [Espinoza, 1955]). Como mencionamos, ese "discurso inédito" tiene interés ya que contiene varias de las ideas que Nirenstein debate en su real o imaginada conversación con Einstein.

Independientemente de que Einstein haya o no pensado en utilizar ese texto como conferencia inaugural de su curso (que es una las posibilidades sugeridas y, de hecho, el discurso comienza con un "Honorable señor Rector, Profesores y Estudiantes de la Universidad") es claro que el texto que Einstein utilizó en su primera conferencia en Buenos Aires es considerablemente más específico y centrado en la evolución de la teoría de la relatividad, antes que en el análisis epistemológico de las ciencias físicas.

**El discurso inédito de Einstein: la versión impresa en *La Vida Literaria***

En ese texto Einstein se felicita de estar en Argentina, "tierra bendita" donde hay también hombres que se interesan por los temas científicos en medio de "luchas económicas y políticas y de subdivisiones nacionalistas". Define a los cultores de la ciencia como separados en dos grandes grupos: el de los que buscan "expandir y enriquecer nuestro saber individual", y el de los que tratan de brindar una "mayor unidad sistemática" al conocimiento.

Si bien se dice que la física es una ciencia empírica, Einstein señala que no existe un método científico que permita pasar de los datos de la experiencia a leyes fundamentales. Einstein percibe a éste último camino como "un ejercicio de la intuición" en el cual se trata de establecer una relación lógica entre los principios y los hechos observados. Destaca que esas leyes fundamentales son provisorias, en cuanto a que un hecho aun no observado puede determinar su cambio. Sin embargo, deja en claro su parecer de que: "La experiencia es, por lo tanto, juez, pero no madre, de las leyes fundamentales". Entre los datos de la experiencia y las leyes fundamentales existe, según este texto, un "acto libre de creación de parte de la fantasía".

En la versión que nos ofrece Nirenstein del pensamiento de Einstein aparece una aclaración no contenida en el texto; dice que "Claro está que solamente tendrá alguna probabilidad de éxito en este empeño el que lo acometa con un dominio empírico suficiente del conjunto de los hechos de que se trate." Y más adelante destaca que "La experiencia juzga, pero no crea las leyes fundamentales." [ambos párrafos de E. sólo aparecen en el texto de M.N.]

Las ideas o conceptos, aunque originados por la experiencia, tienen para Einstein "una cierta independencia lógica." Como ejemplo de que es posible crear conceptos sin una preparación científica hace referencia al surgimiento de la noción de número en las sociedades primitivas.

Cree que no existe tampoco un camino forzoso entre las leyes y los hechos de la experiencia, y da como ejemplo las leyes del movimiento, donde el teorema de Galileo, de que la fuerza es proporcional a la aceleración, "no procede inmediatamente de la experiencia" sino que es una "libre afirmación" que procede del conocimiento adquirido intuitivamente. La historia de la mecánica antes de Galileo nos muestra que esto no es "obvio de por sí". Es un hecho lógicamente arbitrario que la "teoría general de la relatividad ha venido a modific[ar]".

Piensa Einstein que no son sólo las leyes fundamentales las que proceden "de un acto de la fantasía". El concepto mismo de aceleración es para Einstein ejemplo de una "libre creación del espíritu" aunque más tarde se apoye en las estructuras del cálculo infinitesimal.

Señala también que las leyes no sólo pueden ser destruidas a causa de encontrárseles una dependencia errónea o no aplicable en general, sino también "porque los conceptos allí avanzados no se manifiesten de una manera tan clara que se pueda, con su ayuda, dar razón a los hechos observados". A continuación da ejemplos de la historia de la física teórica moderna donde esto ha ocurrido; el concepto elemental de temperatura, frente a un mejor conocimiento de la termodinámica, es uno de ellos.

Einstein recuerda que, concentrándose en volúmenes suficientemente pequeños de un cuerpo, la temperatura fluctúa constantemente alrededor de un valor y las fluctuaciones son tanto mayores cuanto menor es la región bajo estudio. Conecta luego este hecho con la equivalencia entre energía térmica y energía mecánica, para terminar señalando cómo esas fluctuaciones afectan el movimiento de cuerpos pequeñísimos suspendidos en fluidos: el movimiento browniano.

El avance de la ciencia no se da solamente con el desbarranco de teorías por otras más precisas, sino también cuando "las nociones elementales, que corresponden a las realidades fundamentales, deben ser reemplazadas por otras nuevas, más adecuadas al complejo de la experiencia."

Se pregunta Einstein si este desarrollo tiene un término y concluye que no, que "Toda teoría contiene verdad para nosotros solamente en el sentido en que una parábola o una comparación pueden contener la verdad."

Señala finalmente que si bien no podemos así "penetrar hasta las últimas verdades", sin embargo, detrás nuestro vendrá otra generación de investigadores que lograrán llegar más adentro que sus predecesores.

En la versión dialogada de Nirenstein, éste aparece preguntando a Einstein si "Puede señalarse un término a ese desarrollo", a lo que el físico contesta "Los físicos de la actualidad ya no lo creemos. Para nosotros, cualquier teoría contiene tanta verdad como la que puede caber en una ecuación."

### **Primeras conferencias de Einstein en su ciclo de la UBA**

En sus dos conferencias de introducción a su curso sobre la teoría de la relatividad, Einstein utilizó un enfoque diferente del que aparece en el texto inédito: más concentrado en la física, menos en la epistemología de esa disciplina.

En la primera conferencia introductoria, dictada ante la Universidad el Viernes 27 de Marzo [Einstein, 1925b], que sólo tuvo una media hora de duración, Einstein se refirió a los principios fundamentales de la mecánica de Galileo, señalando el rol jugado por los sistemas de referencia en esa formulación, y al papel de las leyes de inercia en ellos. Pasó luego a considerar los experimentos realizados para comprobar la validez del llamado principio de relatividad de la mecánica clásica y la necesidad de realizarlos fuera del campo de la mecánica, para discutir después los experimentos de Fizeau y señalar las dificultades creadas por la noción de espacio absoluto o éter.

En su segunda conferencia introductoria, que fue la primera dictada en la Facultad de Ciencias Exactas el Sábado 28 de Marzo [Einstein, 1925c], Einstein se ocupó en un mayor detalle de los experimentos realizados para verificar la hipótesis de la existencia del éter y las infranqueables dificultades encontradas en esa empresa. Se ocupó luego de las ingeniosas propuestas de Lorentz y FitzGerald para salvar esa hipótesis. Finalmente, hizo referencia a las delicadas experiencias de Michelson que, utilizando métodos de la óptica, permitían detectar con una muy alta precisión fenómenos imposibles de verificar mecánicamente. Esos resultados forzaron a aceptar que las leyes fundamentales de la óptica deben ser independientes de los movimientos de translación uniforme de los sistemas de referencia y, con ello, a generalizar el principio de relatividad de la mecánica, y a aceptar la constancia de la velocidad de la luz.

**Observaciones finales**

Si bien el llamado texto inédito de Einstein no fue utilizado como fundación de su ciclo de conferencias sobre la teoría de la relatividad, Einstein tuvo oportunidad de explayarse sobre tópicos de epistemología de las ciencias físicas en varios otros escenarios.

A principios de Abril y a invitación de Alberini, Einstein dictó la conferencia de apertura de los cursos de la Facultad de Filosofía y Letras. En ésta, el sabio se refirió -en términos generales- a la relación entre los conceptos físicos y la experiencia, y discutió las relaciones entre los principios de la geometría y la realidad.

En el suplemento de la edición de *La Nación* del domingo 12 de Abril de 1925 Alberini publicó una nota sobre la visita de Einstein la Facultad de Filosofía y Letras [Alberini, 1925] en la que dio a publicidad el discurso con el que presentó a Einstein en esa ocasión. Educadamente, Alberini no se anticipa a Einstein en su discurso de introducción, haciendo él mismo referencias específicas a sus teorías; en cambio, destaca el carácter anti-positivista de las ideas de Einstein y hace un ataque velado, aunque bien comprensible para sus contemporáneos, a José Ingenieros.

Es curioso que el discurso inédito de Einstein haya aparecido, inicialmente, extractado y presentado como un diálogo metafórico entre Einstein y Nirenstein (en la nota de 1925 de éste último en la revista *Verbum*).

También llama la atención que se haya demorado seis años en publicarlo y que, cuando finalmente fue traducido y publicado (en 1931), lo fuera en *La Vida Literaria*, una revista literaria de vanguardia de la época.

Es de lamentar que en 1931 (artículo en *La Vida Literaria*), en 1934 (artículo de Espinosa en *La Nación*), o aun en 1955 (publicación *DAVAR*), cuando varios de los principales actores aun estaban en actividad, no se haya dado una respuesta a estos interrogantes, y no se haya hecho un esfuerzo por situar este texto "inédito" con una mayor precisión dentro el cuerpo de actividades de Einstein en la Argentina.

Es interesante destacar que Samuel Glusberg (1898-1987), director de *La Vida Literaria*, y Enrique Espinosa (que veces firma como Enrique Espinoza) son una misma persona. El pseudónimo proviene de una feliz fusión del nombre de Enrique Heine y el apellido del filósofo Benedict (Baruch) de Spinoza.

**Referencias:**

Alberini, Coriolano (1925). Einstein en la Facultad de Filosofía y Letras, *La Nación*, Suplemento, Abril 12.

---------------------- (1980). *Epistolario de Coriolano Alberini*, Tomo I, Diego F. Pró (traducciones, prólogo y notas), Mendoza: Universidad Nacional de Cuyo.

Einstein, Albert (1925a). Pan Europa, *La Prensa*, Marzo 24, 1925.

--------------- (1925b). Einstein dio ayer su primera conferencia en la Universidad, *La Nación*, Marzo 28, 1925.

--------------- (1925c). Einstein expuso sus teorías en la Facultad de Ciencias Exactas, *La Nación*, Marzo 29, 1925.

--------------- (1928). A propos de la déduction relativiste de M. Émile Meyerson, *Revue Philosophique*, 105: 161-166.

--------------- (1931). Un discurso inédito de Alberto Einstein, *La Vida Literaria*, Abril, 30:1.

--------------- (1955). Un discurso inédito de Einstein, *Davar*, 61: 110-113.

Espinosa, Enrique, (1934) El Mundo de Einstein, *La Nación*, 16 de septiembre, segunda sección, 3: 1.

Espinoza, Enrique, (1955). El Mundo de Einstein, "Mein Weltbild", *Davar*, 61: 71-77.

Gangui, Alejandro y Ortiz, Eduardo L. (2005) Crónica de un mes agitado: Albert Einstein visita la Argentina, *Todo es Historia*, 454: 22-30.

Meyerson, Émile (1924). *La déduction relativiste*. Paris: Payot.

Nirenstein, Mauricio (1925), Einstein en Buenos Aires, *Verbum: Revista del Centro de Estudiantes de Filosofía y Letras*, Buenos Aires, XVIII: 167-178.

Ortiz, Eduardo L. (1995). A convergence of interests: Einstein's visit to Argentina in 1925, *Ibero-Amerikanisches Archiv*, Berlín, 20: 67-126.